# Rational Design of High-Tc Superconductivity in the Z = 6.0 Family

O. Paul Isikaku-Ironkwe[1,2]

[1]The Center for Superconductivity Technologies (TCST)
Department of Physics,
Michael Okpara University of Agriculture, Umudike (MOUAU),
Umuahia, Abia State, Nigeria
and
[2]RTS Technologies, San Diego, CA 92121

## Abstract

Rational design of superconductivity from Periodic Table properties is one of the grand challenges of superconductivity. We recently showed (Arxiv: 1208.0071) that high-Tc superconductivity exists in the Z = 5.667 with Ne=2.333. Here we propose and show that materials with Z = 6.0 and Ne =2.0 and 2.22 also meet the conditions for high-Tc superconductivity. We predict that the Ne=2.67 variety will not be superconducting but the ternary and quaternary systems of the Z =6.0 family with Ne=2.0 and 2.22 would have $12.5 \leqq Fw/Z \leqq 25$ and Tcs that fall in the range 60K – 100K. We provide material specific examples of such potential low-Z, Low-Ne high-Tc superconductors.

## Inspiration

"I think we didn't understand superconductivity, in the sense of being able to use the physics to build a molecule with predictable properties. Our understanding was deconstructive or analytical rather than constructive or synthetic. I think there is a difference in these modes of understanding."

--------Roald Hoffmann (August 1990),
Preface to: "Chemistry of Superconductor Materials", Terrell A. Vanderah (Editor)

"Though superconductivity is generally regarded as the domain of physics it is, in fact, intimately associated with synthetic and structural chemistry… "

-------- Jeffrey L. Tallon, The MacDiarmid Institute for Advanced Materials (2011)

Key terms: Rational design of superconductors, material specific computational search

## Introduction

The discovery of MgB$_2$ [1] has opened our eyes to search for superconductors in low-Z, low-Ne materials. Initial searches based on structural and electronic symmetry with MgB$_2$ did not yield Tcs near that of MgB$_2$ [2 - 8]. Guided by some simple empirical rules [9] we initiated searches with a novel strategy based on atomic number, Z, and valence electrons, Ne, and electronegativity, $x$, symmetry. Inspired by the MgB$_2$ paradigm of high-Tc superconductivity in low-Z, low-Ne materials, we applied the discoveries from reference [9] to search beyond ternary low-Z materials into the new world of quaternary MgB$_2$-like superconductors. Pickett, in a thought-provoking paper [10] had asked the "perverse" question: *"Why is Tc of MgB$_2$ only 40K? Why isn't it 60K or 100K?"* We answer that question in this paper. In our premier paper [9] we established new criteria for achieving higher-Tc superconductivity: increase the Fw/Z by increasing the number of elements, En, and atoms, An, while keeping Ne/SqrtZ between 0.8 and 1.0. Using this strategy, we pursued the search for MgB$_2$-like superconductors beyond the ternary level and beyond the Fw/Z =6.27 limit value for MgB$_2$. Recently, we predicted new MgB$_2$-like superconductors [11 - 16] and new atomic-number families (Fz) [17 – 22] that should harbor high-Tc materials. In this paper we explore the spectrum of possible combinatorial synthesis of families ranging from Z=2.0 to Z =10.0 that could produce the Z=6.0 family. We propose that the Z=6.0 family should produce a rich treasure house of quaternary 6-atom systems with Ne= 2.0, 2.22 and 2.67 many of which should be high-Tc superconductors. This paper is structured as follows: We first present the fundamental framework of our search strategy. Next we show combinatorial synthesis of families of materials that should yield Z=6.0 family with Ne = 2.0, 2.22 and 2.67. We then proceed to show that some will be superconducting and others not. We next estimate the possible Tcs of the Z=6.0 materials predicted to be superconductors. We discuss too the concept of rational design of superconductors and the material specific computational search for high-Tc superconductivity.

**Framework for Searching for Superconductivity**

Tc(max) can be expressed in terms of the Periodic table properties of electronegativity, $x$, valence electron count per atom, Ne, atomic number, Z and formula weight, Fw, and is derived in [9] as:

$$T_c = x \frac{Ne}{\sqrt{Z}} K_o \tag{1}$$

where Ko = n(Fw/Z) and n can be determined empirically(see fig. 6). For magnesium diboride, we found that n = 3.65. For other superconducting materials, n varies between 1 and 3.65 [9].

For a material with formula: $A_pB_qC_rD_sE_t$, the average values of $\mathcal{X}$, Ne and Z are computed as:

$$\mathcal{X} = \frac{p\mathcal{X}_A + q\mathcal{X}_B + r\mathcal{X}_C + s\mathcal{X}_D + t\mathcal{X}_E}{p+q+r+s+t} \qquad (2)$$

$$Ne = \frac{pNe_A + qNe_B + rNe_C + sNe_D + tNe_E}{p+q+r+s+t} \qquad (3)$$

$$Z = \frac{pZ_A + qZ_B + rZ_C + sZ_D + tZ_E}{p+q+r+s+t} \qquad (4)$$

The formula weight, Fw, for a material $A_pB_qC_rD_sE_t$ is the sum of the weighted components and is given as:

$$Fw = pFw_A + qFw_B + rFw_C + sFw_D + tFw_E \qquad (5)$$

With these data we can assembly the material specific characterization dataset (MSCD) of the material and evaluate it for superconductivity and estimate its Tc.

## Tc estimation from material specific equations

After superconductivity is established using equation (13), Tc may be estimated with equation (1) above. Fw/Z can be computed from equations (4) and (5) and the terms of equation (1) from equations (2), (3) and (4). The graph of Tc versus Fw/Z in figure 6 helps us to evaluate the numerical value of n and estimate Tc from the Ko = (Fw/Z)n value. Usually, n lies between 1 and 3.65, the value for $MgB_2$, after we have estimated the occurrence of high-Tc superconductivity, as shown in reference [9] and expressed below:

$$0.8 < Ne/\sqrt{Z} < 1.0 \qquad (6)$$

## Combinatorial Synthesis for Z = 6.0 Family

Synthesizing the $F_{z=6.0}$ family of materials may be achieved as shown in equations (7) to (9), illustrated in figures 1 to 3 and displayed in Tables 6 to 8. The synthesis equations are:

$$F_{Z=4.67} + F_{Z=7.33} = F_{Z=6.0} \quad \{Ne=2.0, 2.67\} \qquad (7)$$

$$F_{Z=2.0} + F_{Z=10.0} = F_{Z=6.0} \quad \{Ne=2.0\} \qquad (8)$$

$$2F_{Z=4.67} + F_{Z=8.67} = F_{Z=6.0} \quad \{Ne=2.22, 2.67\} \qquad (9)$$

The material specific characteristics dataset (MSCD) of the $F_{Z=6.0}$ family of materials will be:

$$\text{MSCD of Material} = \langle \mathcal{X}, \text{Ne}, Z, \text{Ne}/\sqrt{Z}, \text{Fw}, \text{Fw}/Z \rangle \tag{10}$$

$$= \langle \mathcal{X}, 2.0, 6.0, 0.8165, \text{Fw}, \text{Fw}/Z \rangle \tag{11}$$

$$= \langle \mathcal{X}, 2.22, 6.0, 0.9071, \text{Fw}, \text{Fw}/Z \rangle \tag{12}$$

$$= \langle \mathcal{X}, 2.67, 6.0, 1.0888, \text{Fw}, \text{Fw}/Z \rangle \tag{13}$$

Here one of the materials has Ne = 2.0 and the others Ne= 2.22 and 2.667. Thus we should expect high-Tc in the Ne=2.0 and Ne=2.22 materials of equations (11) and (12) from the relation of (6).

## Results

1. $F_{Z=4.67} + F_{Z=7.33} = F_{Z=6.0}$ {Ne=2.0, 2.67}

Three methods were used to design Z=6.0 materials. The first in (7) above involves combinatorial synthesis of $F_{Z=4.67}$ and $F_{Z=7.33}$ materials. The MSCD of $F_{Z=4.67}$ family in Table 1 has two classes of materials: those with Ne=2.67 and those with Ne=1.33. For the $F_{Z=7.33}$ family we choose $CaH_2$ (Table 2) which has Ne=1.33. We react it with $F_{Z=4.67}$ family with Ne=2.67 to get $F_{Z=6.0}$ family with Ne=2.0 as shown in figure 1. The potential products are shown in Table 6, with $12.0 \leqq Fw/Z \leqq 20.0$

2. $F_{Z=2.0} + F_{Z=10.0} = F_{Z=6.0}$ {Ne=2.0}

The second method involves the combinatorial synthesis of $F_{Z=2.0}$ and $F_{Z=10.0}$ materials. The MSCD of $F_{Z=2.0}$ material ($BeH_2$) is given in Table 3 with Ne=1.33. The MSCD of $F_{Z=10.0}$ family is shown in Table 4 which has Ne=2.67. Combinatorial synthesis of the two families should yield the $F_{Z=6.0}$ family with Ne=2.0 as shown in figure 2. The potential products are shown in Table 7, with $18 \leqq Fw/Z \leqq 20.0$.

3. $2F_{Z=4.67} + F_{Z=8.67} = F_{Z=6.0}$ {Ne=2.22, 2.67}

The third method involves the combinatorial synthesis of two parts of $F_{Z=4.67}$ material with Ne=2.67 with a $F_{Z=8.67}$ material with Ne=1.33. The MSCD of $F_{Z=4.67}$ materials is given in Table 1. The MSCD of $F_{Z=8.670}$ family is shown in Table 5 with products in Table 8. Here we see that a Fw/Z of 19 can be attained with electronegativity peaking at 1.589.

**Discussion**

"Rational design of superconductors" has been and still is a major challenge in the field of superconductivity. It all starts with the chemical building blocks of superconductors (the elements of the Periodic Table) and predicting the occurrence of superconductivity, then estimating Tc in terms of correlated properties of the elements(electronegativity, valence electrons, atomic number and weight). The BCS theory [23] explained the microscopic physics of superconductivity but has failed to provide the macroscopic chemical building blocks of superconductivity [24, 25]. The DOE-BES report [24] on superconductivity recommended that: *"A theory with some degree of material-specificity is a pre-requisite for it to be useful in guiding a search for better superconductors…"* In our paper [9] we through a semi-empirical and phenomenological approach, derived a chemical relation of Tc maximum in terms of correlations of superconductivity with electronegativity, valence electrons and atomic number and formula weight of the material under consideration. The occurrence of high-Tc is given in chemical terms as: $0.8 < Ne/\sqrt{Z} < 1.0$ and Tc (maximum) as: $T_c = \mathcal{X} \frac{Ne}{\sqrt{Z}} K_o$ with terms explained above and in reference [9]. Working with these two material specific derivations and the concept of material specific characterization datasets (MSCD) also derived in [9], we have been able to venture into the hitherto almost forbidden territory of "rational design of superconductors". It is well known that serendipity has dominated discoveries in superconductivity [24, 26]. This has been so because physicists, chemists and computer scientists have not really collaborated sufficiently to address the chemistry, physics and computer science involved in the "material specific computational search" for high-Tc superconductivity. This paper and previous ones [9, 11 - 22] point the way to true rational design of superconductivity from a chemical and computational perspective direction based on the Periodic Table properties of superconductivity and search algorithms.

Searching for superconductivity in the Z=6.0 family reveals a rich factory of potential compounds whose thermodynamic stability need to be further investigated experimentally. The computational combinatorial method herein used makes it possible to look at the spectrum of possibilities that call for further laboratory exploration. The Z=6.0 family allows for values of Ne=2.0. 2.22 and 2.67. The combinatorial families that resulted in Z=6.0 leads us to observe a sequence that may be approximated as: Z =1.333p + 2 where $0 \leqq p \leqq 8$. The Z-sequence is: 2, 3.333, 4.667, 6.0, 7.333, 8.667, 10.0, 11.333, 12.667. The only members yet to be explored for superconductivity are the Z=2.0, Z=3.33 and Z=11.33. The method described here does not explore in depth the potential phases that may form. These are best dealt with using DFT methods [27, 28, 29] or by direct experiment. Our method has revealed, for the first time that quaternary $MgB_2$-like materials with $12.5 \leqq Fw/Z \leqq 25$ do exist and with Tcs between 60K and 100K, answering the "perverse" question earlier posed by Pickett [10]. Five families (fig.4) were involved in creating Z=6.0. The possible number of combinations to produce the Z=6.0 run well over 100. Thus the Z =6.0 is a rich treasure trove for low-Z, low-ne high-Tc superconductivity.

## Conclusion

Combinatorial synthesis of low-Z families with Z = 2.0, 4.667, 7.333, 8.667 and 10.0 can lead to a new family (Z=6.0) of ternary, quaternary and 5-element materials. The many possible Z =6.0 materials with Ne=2.0, 2.22 are predicted to have $12.5 \leqq Fw/Z \leqq 25$ and Tcs ranging from 60K up to 100K for the quaternary and 5-element types. Z=6.0 materials with Ne=2.667 are predicted to be non-superconducting. In the computational investigation of the Z=6.0 family, a new quaternary family, Z=5.56 was discovered with Fw/Z=18.4 with estimated Tc of 105K. The Z=6.0 family joins the other families (Z = 1.333p +2, where $0 \leqq p \leqq 8$) that have been designed [17 – 22] using our unique material specific design methodology (MSDM) The experimental verifications of these predictions are awaited.

## Acknowledgements

This research was facilitated through generous grants from my mentor, M. J. Schaffer, formerly at General Atomics, San Diego. Many discussions with A.O.E. Animalu at the University of Nigeria, Nsukka, and with M. Brian Maple at University of California, San Diego and J.R. O'Brien at Quantum Design, San Diego proved stimulating useful and continuously challenging in the search for rational design of superconductors.

**Figures**

**Tables**

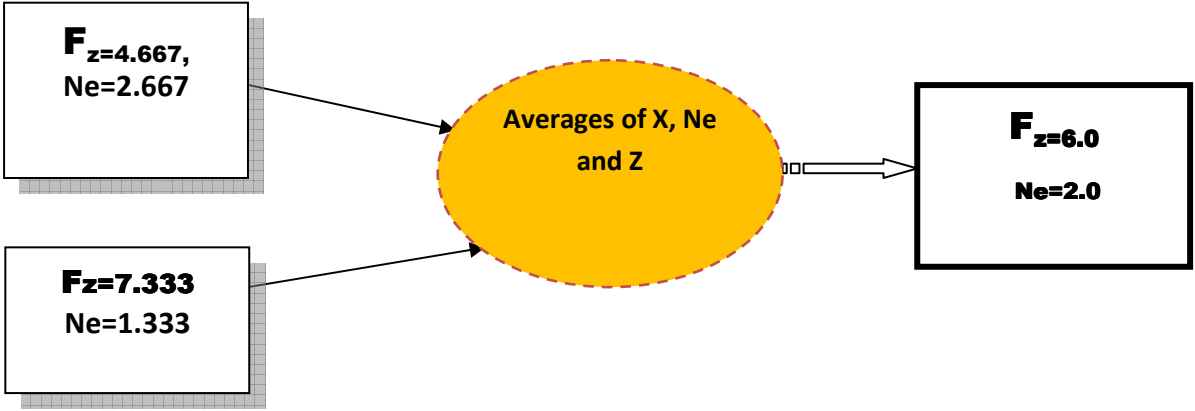

**Figure1:** Combinatorial synthesis of Fz=6.0 family of materials from Fz=4.667 and Fz=7.333 materials. Examples of this combination are shown in Table 6.

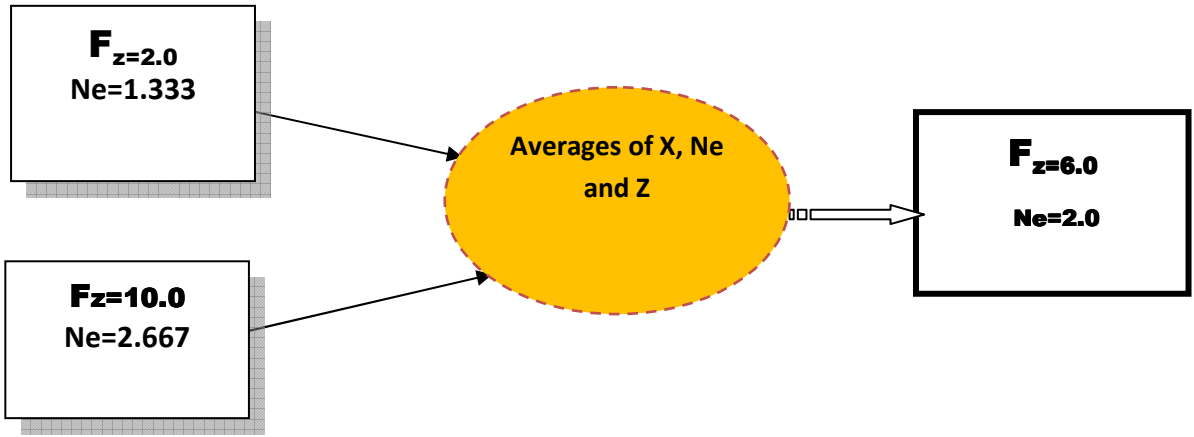

**Figure 2:** Combinatorial synthesis of Fz=2.0 family of materials and Fz=10.0 yields Fz=6.0. Examples of this combination are shown in Table 7.

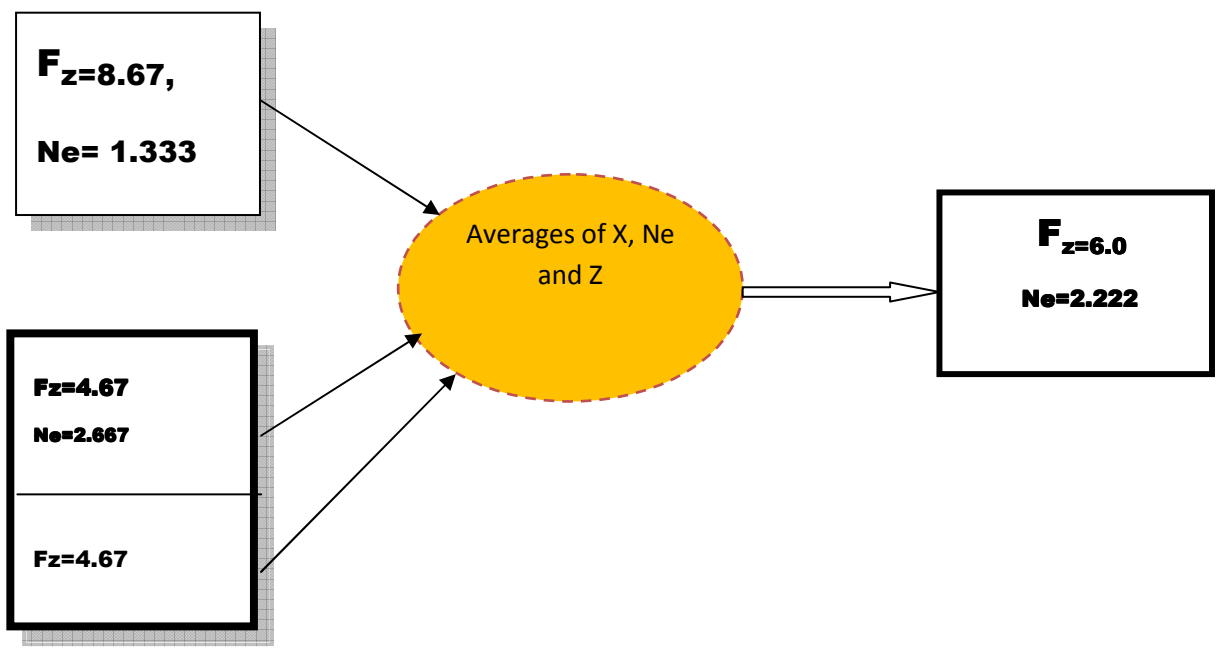

**Figure 3:** Combinatorial synthesis of Fz=6.0 family of materials from Fz=8.67 with Ne=1.333 and TWO Fz=4.67 materials with Ne=2.667. Examples of this combination are shown in Table 8.

| | Combinatorics | Example | Z | $x$ | Ne/$\sqrt{Z}$ | An | En | n | Fw/Z | Tc range |
|---|---|---|---|---|---|---|---|---|---|---|
| 1 | $F_{Z=4.67} + F_{Z=7.33}$ | Be$_2$CaCH$_2$ | 6.0 | 1.783 | **0.8165** | 6 | 4 | 3.66 | 12.02 | 64K |
| 2 | $F_{Z=2.0} + F_{Z=10.0}$ | NaBeBSiH$_2$ | 6.0 | 1.733 | **0.8165** | 6 | 5 | 3.72 | 12.15 | 64K |
| 3 | $2F_{Z=4.67} + F_{Z=8.67}$ | Li$_2$Na$_2$BeB$_2$C$_2$ | 6.0 | 1.589 | **0.9071** | 9 | 5 | 3.36 | 19.1 | 92.5 |
| 4 | $2F_{Z=4.67} + F_{Z=7.33}$ | Be$_4$CaC$_2$H$_2$ | 5.56 | 1.8 | **0.9427** | 9 | 4 | 3.36 | 18.4 | 105K |

**Figure 4:** Combinatorial synthesis paths to Fz=6.0 and Fz=5.56 key results. An=# of atoms; En=# of elements, n=number empirically determined from graph of Tc vs Fw/Z (Figure 6). Tcs are then estimated using equation (1).

| Fz = | 2.0 | 4.67 | 4.67 | 7.33 | 8.67 | 10.0 |
|------|-----|------|------|------|------|------|
| 2.0  |     |      |      |      |      | 🟦   |
| 4.67 |     |      |      |      | 🟥   |      |
| 4.67 |     |      |      | 🟩   | 🟥   |      |
| 7.33 |     |      | 🟩   |      |      |      |
| 8.67 |     | 🟥   | 🟥   |      |      |      |
| 10.0 | 🟦  |      |      |      |      |      |

**Figure 5:** Combinatorial synthesis families for producing Fz=6.0, depicted in colors. Blue represents Fz=2.0 and Fz=10.0; Brown represents 2(Fz=4.67) and Fz=8.67; Green represents Fz=4.67 and Fz=7.33.

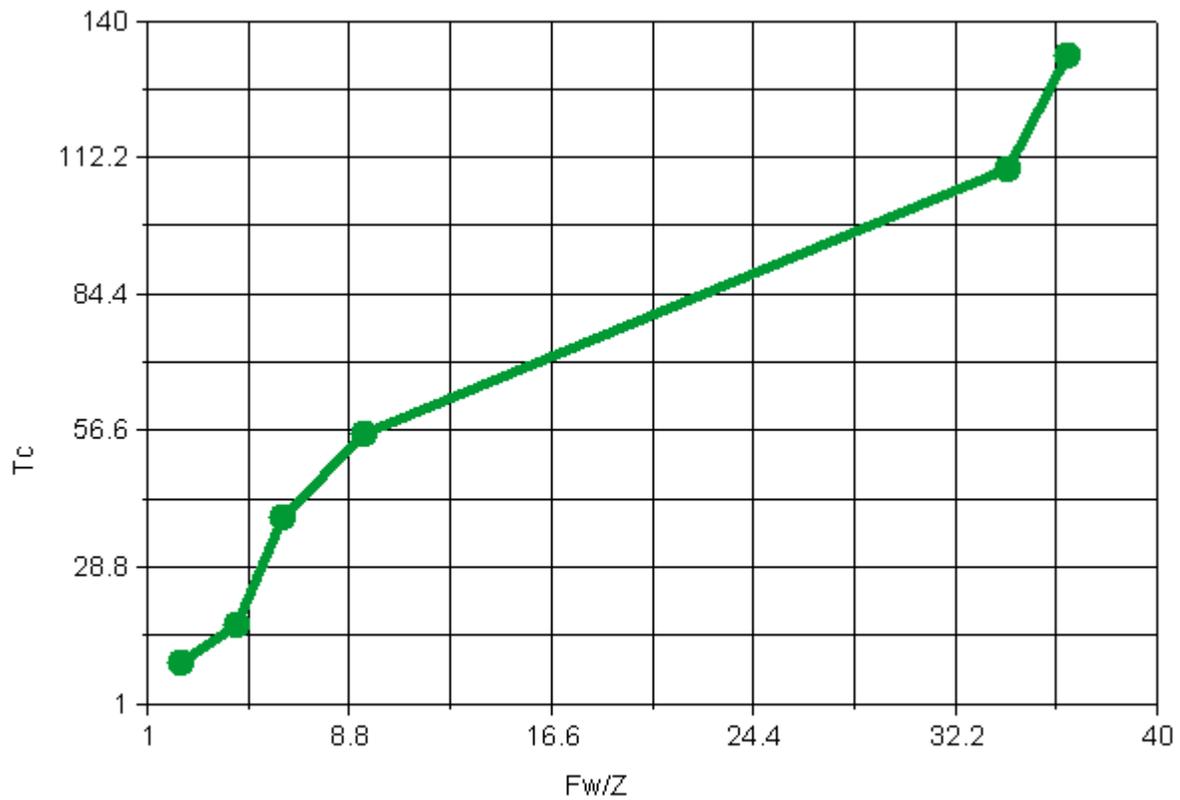

Tc vsFw/Z for 6 Families of Superconductors. The graph indicates that as Fw/Z increases, Tc increases too.

**Figure 6:** Adapted from Reference [9] , this graph helps us to estimate Tc knowing Fw/Z and thus to estimate n in the formula for Tc.(equation 1). Note that for MgB$_2$, with Fw/Z =6.27, n =3.65.

| | Material | $x$ | Ne | Z | Ne/$\sqrt{Z}$ | Fw | Fw/Z |
|---|---|---|---|---|---|---|---|
| 1 | LiBC | 1.8333 | 2.6667 | 4.6667 | 1.2344 | 29.76 | 6.377 |
| 2 | BeB$_2$ | 1.8333 | 2.6667 | 4.6667 | 1.2344 | 30.63 | 6.564 |
| 3 | Be$_2$C | 1.8333 | 2.6667 | 4.6667 | 1.2344 | 30.03 | 6.435 |
| 4 | Li$_2$O | 1.8333 | 2.6667 | 4.6667 | 1.2344 | 29.88 | 6.403 |
| 5 | LiBeN | 1.8333 | 2.6667 | 4.6667 | 1.2344 | 29.96 | 6.420 |
| 6 | LiB$_5$ | 1.8333 | 2.6667 | 4.6667 | 1.2344 | 60.99 | 13.07 |
| 7 | Li$_3$BN$_2$ | 1.8333 | 2.6667 | 4.6667 | 1.2344 | 59.65 | 12.78 |
| 8 | NaAlH$_4$ | 1.8 | 1.3333 | 4.6667 | 1.2344 | 54.01 | 11.574 |
| 9 | KBH$_4$ | 1.8667 | 1.3333 | 4.6667 | 1.2344 | 53.95 | 11.56 |
| 10 | MgH$_2$ | 1.8 | 1.3333 | 4.6667 | 1.2344 | 26.33 | 6.642 |

**Table 1:** 7 MSCDs of Z =4.67 materials with Ne=2.67 and 3 with Ne=1.333.

| Material | $x$ | Ne | Z | Ne/$\sqrt{Z}$ | Fw | Fw/Z |
|---|---|---|---|---|---|---|
| CaH$_2$ | 1.7333 | 1.333 | 7.333 | 0.4923 | 42.1 | 5.74 |

**Table 2:** MSCD of Z =7.333 material (CaH$_2$) with Ne= 1.333.

| | Material | $x$ | Ne | Z | Ne/$\sqrt{Z}$ | Fw | Fw/Z |
|---|---|---|---|---|---|---|---|
| 1 | BeH$_2$ | 1.9 | 1.333 | 2.0 | 0.9426 | 11.03 | 5.515 |

**Table 3:** MSCD of Z =2.0 material with Ne= 1.333.

| | Material | $\mathcal{X}$ | Ne | Z | Ne/$\sqrt{Z}$ | Fw | Fw/Z |
|---|---|---|---|---|---|---|---|
| 1 | MgBeSi | 1.833 | 2.667 | 10.0 | 0.8433 | 61.41 | 6.141 |
| 2 | Na$_2$O | 1.767 | 2.667 | 10.0 | 0.8433 | 61.98 | 6.198 |
| 3 | KBeN | 1.767 | 2.667 | 10.0 | 0.8433 | 62.12 | 6.212 |
| 4 | NaMgN | 1.7 | 2.667 | 10.0 | 0.8433 | 61.31 | 6.131 |
| 5 | LiCaN | 1.667 | 2.667 | 10.0 | 0.8433 | 61.03 | 6.103 |
| 6 | CaB$_2$ | 1.667 | 2.667 | 10.0 | 0.8433 | 61.70 | 6.170 |
| 7 | CaBeC | 1.667 | 2.667 | 10.0 | 0.8433 | 61.10 | 6.110 |
| 8 | Mg$_2$C | 1.633 | 2.667 | 10.0 | 0.8433 | 60.63 | 6.063 |
| 9 | NaAlC | 1.633 | 2.667 | 10.0 | 0.8433 | 61.98 | 6.198 |
| 10 | BeBSc | 1.6 | 2.667 | 10.0 | 0.8433 | 64.78 | 6.478 |
| 11 | NaBSi | 1.567 | 2.667 | 10.0 | 0.8433 | 61.89 | 6.189 |
| 12 | BeAl$_2$ | 1.5 | 2.667 | 10.0 | 0.8433 | 62.97 | 6.297 |
| 13 | Be$_2$Ti | 1.5 | 2.667 | 10.0 | 0.8433 | 65.90 | 6.590 |
| 14 | NaBeP | 1.5 | 2.667 | 10.0 | 0.8433 | 62.97 | 6.297 |
| 15 | LiNaS | 1.467 | 2.667 | 10.0 | 0.8433 | 62.00 | 6.200 |
| 16 | LiMgP | 1.433 | 2.667 | 10.0 | 0.8433 | 62.22 | 6.222 |

**Table 4:** 7 MSCDs of Z =10.0 materials with Ne=2.67.

| | Material | $\mathcal{X}$ | Ne | Z | Ne/$\sqrt{Z}$ | Fw | Fw/Z |
|---|---|---|---|---|---|---|---|
| 1 | CaLi$_2$ | 1.0 | 1.333 | 8.667 | 0.4528 | 53.96 | 6.226 |
| 2 | BeNa$_2$ | 1.1 | 1.333 | 8.667 | 0.4528 | 54.99 | 6.345 |

**Table 5 :** MSCD of Z =8.667 materials with Ne= 1.333.

| Materials (Z=7.33 + Z=4.67) & Z=7.33 + 2(Z=4.67) | $\chi$ | Ne | Z | Ne/$\sqrt{Z}$ | Fw | Fw/Z |
|---|---|---|---|---|---|---|
| 1 | $CaH_2$ + LiBC = $LiCaBCH_2$ | 1.783 | 2.0 | 6.0 | 0.8165 | 71.86 | 11.97 |
| 2 | $CaH_2$ + $Be_2C$ = $Be_2CaCH_2$ | 1.783 | 2.0 | 6.0 | 0.8165 | 72.13 | 12.022 |
| 3 | $CaH_2$ + $BeB_2$ = $BeCaB_2H_2$ | 1.783 | 2.0 | 6.0 | 0.8165 | 72.73 | 12.122 |
| 4 | $CaH_2$ + LiBeN = $LiCaBeNH_2$ | 1.7833 | 2.0 | 6.0 | 0.8165 | 72.06 | 12.01 |
| 5 | $CaH_2$ + $Li_3BN_2$ = $Li_3CaBN_2H_2$ | 1.8 | 2.222 | 5.556 | 0.9427 | 101.75 | 18.32 |
| 6 | $CaH_2$ + $LiB_5$ = $LiCaB_5H_2$ | 1.8 | 2.222 | 5.556 | 0.9427 | 103.09 | 18.56 |
| 7 | $CaH_2$ + 2(LiBC) = $CaLi_2B_2C_2H_2$ | 1.8 | 2.222 | 5.556 | 0.9427 | 101.62 | 18.29 |
| 8 | $CaH_2$ + 2($Be_2C$) = $Be_4CaC_2H_2$ | 1.8 | 2.222 | 5.556 | 0.9427 | 102.16 | 18.39 |

**Table 6:** MSCD of materials obtained by combinatorial synthesis of Z =7.333 material ($CaH_2$) with Ne= 1.333 and Z=4.667 material with Ne=2.667. Note that using two times the Z=4.67 material leads to a new family Z=556 with higher Fw/Z=18.4 whose Tc is estimated at 105K. The Fw/Z =12.0 materials have Tc estimated at 63.6K.

| Materials (Z=2.0 + Z=10.0) | $\chi$ | Ne | Z | Ne/$\sqrt{Z}$ | Fw | Fw/Z |
|---|---|---|---|---|---|---|
| 1 | $BeH_2$ + $Na_2O$ = $Na_2BeH_2O$ | 1.833 | 2.0 | 6.0 | 0.8165 | 73.01 | 12.17 |
| 2 | $BeH_2$ + $BeAl_2$ = 2BeAlH | 1.7 | 2.0 | 6.0 | 0.8165 | 37.0 | 6.17 |
| 3 | $BeH_2$ + $Mg_2C$ = $BeMg_2H_2C$ | 1.567 | 2.0 | 6.0 | 0.8165 | 71.66 | 11.94 |
| 4 | $BeH_2$ + LiMgP = $LiBeMgPH_2$ | 1.667 | 2.0 | 6.0 | 0.8165 | 73.25 | 12.21 |
| 5 | $BeH_2$ + NaBeP = $NaBe_2PH_2$ | 1.7 | 2.0 | 6.0 | 0.8165 | 74.0 | 12.33 |
| 6 | $BeH_2$ + CaBeC = $Be_2CaCH_2$ | 1.783 | 2.0 | 6.0 | 0.8165 | 72.13 | 12.02 |
| 7 | $BeH_2$ + NaBSi = $NaBeBSiH_2$ | 1.733 | 2.0 | 6.0 | 0.8165 | 72.92 | 12.15 |

**Table 7:** MSCD of materials obtained by combinatorial synthesis of Z =2.0 materials with Ne= 1.333 and Z=10.0 materials with Ne=2.667 to give Z=6.0. The Fw/Z for the binary compound is 6.2 and 12.15 for the quaternary compound. Ne/$\sqrt{Z}$ =0.8165 implies that the material is likely a superconductor with Tc estimated at 64K.

.

| Materials : Z=8.67 + 2(Z=4.67) | $\mathcal{X}$ | Ne | Z | Ne/$\sqrt{Z}$ | Fw | Fw/Z |
|---|---|---|---|---|---|---|
| 1 | CaLi$_2$ + 2(LiBC) = Li$_4$CaB$_2$C$_2$ | 1.556 | 2.222 | 6.0 | 0.9071 | 113.48 | 18.91 |
| 2 | BeNa$_2$ +2(LiBC) = Li$_2$Na$_2$Be B$_2$ C$_2$ | 1.589 | 2.222 | 6.0 | 0.9071 | 114.51 | 19.09 |
| 3 | CaLi$_2$ + 2(Be$_2$C) = Li$_2$Be$_4$CaC$_2$ | 1.556 | 2.222 | 6.0 | 0.9071 | 114.02 | 19.00 |
| 4 | BeNa$_2$ +2(Be$_2$C) = Na$_2$Be$_5$C$_2$ | 1.589 | 2.222 | 6.0 | 0.9071 | 115.05 | 19.18 |
| 5 | CaLi$_2$ +2(Li$_2$O) = Li$_6$CaO$_2$ | 1.556 | 2.222 | 6.0 | 0.9071 | 113.72 | 18.95 |

**Table 8:** MSCD of materials obtained by combinatorial synthesis of Z =8.667 materials with Ne= 1.333 (Table 4) and Z=4.667 materials with Ne=2.667 (Table 1) to give Z=6.0. The Fw/Z for the ternary and quaternary compounds is between18.91 and 19.1 for the quaternary compound. Ne/$\sqrt{Z}$ =0.9071 implies that the material is likely a superconductor with Tc estimated at 93K.

**COMMENTS on this Paper………?**